\documentclass{iopart}
\begin{document}
\jl{6}
\title{How much  will we learn from the CMB ?}
\author{
David Langlois\footnote[1]{E-mail address: {\tt langlois@iap.fr}}}

\address{Institut d'Astrophysique de Paris, \\
 Centre National de la Recherche Scientifique (UPR 341), \\
98bis Boulevard Arago, 75014 Paris, France}
\date{\today}
\begin{abstract}
The purpose of this article is to give a brief account of what 
we hope to learn from the future CMB experiments, essentially from the 
point of view of primordial cosmology. After recalling what we have 
already learnt,  the principles of parameter extraction from
the data are summarized. The discussion is then devoted to the 
information we could gain about the early universe, in the framework 
of the inflationary scenario, or in more exotic scenarios like brane 
cosmology.  
\end{abstract}

\def\beq{\begin{equation}}
\def\eeq{\end{equation}}

\section{What we have learnt already}
I will start  by recalling briefly what we have already learnt
from the Cosmic Microwave Background (CMB), 
which is already in itself rather impressive. 
For more details, the reader is invited to refer to the numerous reviews 
on the subject, for example the lecture notes of the 1993 les Houches 
school by R. Bond \cite{bond96} and those of the 1999 les Houches 
school by F. Bouchet, J.L. Puget and J.M. Lamarre \cite{bouchet99}.

As is well known, the CMB was predicted in 1948 \cite{cmb48} and discovered 
less than twenty years later \cite{cmb65}. 
Since then, it has been measured repeatedly, and with 
increasing precision.
The CMB spectrum is  that of 
a black body to a very high precision. This feature  gives  
one of the strongest 
arguments in favour of the hot big bang model, according to which 
 the photons where in thermal equilibrium in the past.
This also implies some very stringent constraints on energy release in the 
universe after a redshift $z\sim 10^{6-7}$.

A second important feature of the CMB is that it is almost isotropic but
not quite. 
There is first a dipole at the $10^{-3}$ level, which is usually interpreted 
as the motion of Earth with respect to the CMB rest frame. 
There are then higher multipole anisotropies at the $10^{-5}$ level, 
which had been expected for a long time, and observed for the first time 
by the COBE satellite ten years ago \cite{cobe92}. 
These anisotropies had been 
expected for a long time because they were believed (and 
are still today) to be generated at 
the moment of last scattering by   
very tiny cosmological fluctuations,  the ancestors of 
the present cosmological  structures.

Before going on, let us recall quickly the basic formalism to 
describe the CMB anisotropies. The temperature anisotropies 
can be expanded in terms of (scalar) spherical harmonics,
\beq
{\Delta T\over T}(\theta,\phi)=\sum_{l,m} a_{lm} Y_{lm}(\theta, \phi).
\label{alm}
\eeq
For the theorist, the temperature is an isotropic random field and therefore
the multipole coefficients $a_{lm}$ random variables. One can define 
the angular power spectrum by 
\beq
C_l=\langle |a_{lm}|^2\rangle,
\label{Cl}
\eeq
which is enough to specify entirely  the temperature random field 
if it is Gaussian.

What is usually plotted is the quantity $l(l+1) C_l$.  
If one assumes coherent scale-invariant initial fluctuations, such 
as those produced by inflation,  one  
expects to observe a plateau at low $l$, corresponding to large 
angular scales. At smaller scales, one expects to see oscillations 
\cite{oscillations}.
The reason is that a given  Fourier mode, characterized by a constant 
comoving wavenumber $k$, will start to oscillate as soon as its 
wavelength is within the Hubble radius, i.e. when pressure enters into 
play. Of course, different Fourier modes enter the Hubble radius at 
different times and thus, at a given time, they are at different stages
of their oscillatory pattern. 
The CMB, being then essentially a snapshot of the last scattering surface, 
we thus expect to see oscillations in the angular power spectrum.

By contrast, the   topological defect models, the main competitor
facing  inflation for many years, do not predict oscillations because
the contributions of many incoherent fluctuations, generated at different 
times by topological defects, add and smear out the oscillations 
\cite{defects}.

With the recent data 
from Boomerang \cite{boomerang}, Maxima \cite{maxima}
 and DASI \cite{dasi}, the picture which is now emerging 
is that predicted by inflation and not by topological defects, which  
cannot account for  the main part of the initial spectrum.
Moreover  the position of the first peak 
suggests  that our Universe is quasi-flat.

In less than ten years, we have thus learnt a lot from the CMB anisotropies.
The  coming decade should be extremely fruitful as well, with several 
planned experiments, the most ambitious being the Planck 
satellite mission \cite{planck}.
So far, it is however remarkable, for cosmology, that there has been
no real surprise (apart a quantitative surprise with the amplitude of the 
anisotropies)  with the CMB. The simplest theoretical models 
were  able to  predict in 
advance what we have observed.
The question is how long this situation 
 is going to last. With the increasing precision
of the forthcoming experiments, will the simplest early universe models 
survive or will one need slightly more complicated models, a lot of 
which have already been explored by theorists ?

\section{What we hope to learn about the cosmological parameters}
Before discussing the extraction of cosmological parameters from the CMB
data, it is important to recall that the actual CMB signal is the sum 
of three components:
\begin{itemize}
\item primary anisotropies
\item secondary anisotropies
\item astrophysical foregrounds.
\end{itemize}
The separation of these three components will involve a lot of work 
in data analysis as well as some understanding of the physical processes 
producing these components.
Although in this review I will focus on the early universe perspective, 
for wich the relevant information comes from the primary anisotropies, 
it is worth emphasizing that the  other components also provide 
useful information in various fields of cosmology and 
astrophysics. For instance,
the Sunyaev-Zeldovich effect corresponding to the scattering of CMB photons
from hot gas in galaxy clusters, provides an important proble of the 
physics of clusters.

Let us now concentrate on the primary anisotropies. For a scale-invariant
initial spectrum, the expected spectrum today is composed of 
a plateau at small $l$ and of a succession of 
acoustic peaks at larger $l$, as mentioned before. 
What is important for cosmology is that  
the height and angular position 
 of the acoustic peaks are directly related to 
the  parameters of the cosmological model. In fact  it makes more sense 
  to divide this set of parameters into two classes: 
\begin{itemize}
\item the parameters that define the  geometry and matter content of 
the universe: this in general 
includes the Hubble parameter $H$, the spatial 
curvature $\Omega_k$, the total energy density $\Omega$, the baryonic
contribution $\Omega_b$, the cosmological constant $\Omega_\Lambda$
(this could be replaced by a time-dependent quintessence type matter 
component), 
and the cold dark matter contribution $\Omega_c$, where we have used 
the generic notation $\Omega_X=8\pi G\rho_X/3H^2$ for any component 
with energy density $\rho_X$;

\item the parameters that define the ``initial'' fluctuation spectra,
i.e. the spectra produced during the early universe: this usually 
includes the normalisation of the scalar spectrum, 
its  index $n_s$, the tensor spectrum index $n_t$ and the tensor/scalar 
amplitude ratio.
\end{itemize}
There is a difference of nature between these two sets of 
parameters in the usual framework of perturbed FLRW cosmology. 
The first set of parameters is associated with the homogenous part 
of cosmology (there might be some unknown in the total 
number of components of dark matter) whereas the second set of parameters
is aimed at parametrizing  spectra, i.e.  functions. 
It is preferable to reserve the name of `cosmological parameters' to 
only the first set and to denote the second set by `initial perturbation'
parameters to insist on their different nature.

One of the  present quest of  cosmology is to determine, with the highest 
possible precision, the value of the cosmological parameters (in the 
restrictive sense)  from the CMB data to come. 
One must be aware that the ``measurement'' of these cosmological 
parameters, as well as the expected precision, can 
depend sensitively on the assumptions concerning the initial 
perturbation spectra and their parametrization. Reversing the perspective, 
it will be indispensable  to combine CMB data with other measurements 
of the cosmological parameters in order to get as much information as possible 
on the initial perturbation spectra.

It is now instructive to recall the principle of the extraction of 
 the cosmological parameters from the CMB data. 
Let us start from a CMB map (which assumes a lot of work in data analysis 
has already been done), which we call 'D' (for data). Let us call the set 
of cosmological parameters we wish to extract 'T' (for theory).

The idea is to use Baye's theorem, which simply expresses in two ways 
the joint probability to have both D and T:
\beq
 P(T|D) P(D)= P(D|T) P(T),
\eeq
where $P(D|T)$ is usually called the likelihood and denoted $L(D|T)$.
The method is then to maximize the likelihood function, given the data, 
and thus to obtain some estimates $\hat T (D)$ of the parameters. 
Expanding ${\cal L}=-\ln L$ in the neighbourhood of $\hat T$, $L$ can 
be approximated by a Gaussian distribution. The 
expectation value of the inverse covariance matrix 
\beq
F_{ij}=\langle {\partial^2{\cal L}\over\partial T_i\partial T_j}
\rangle,
\eeq
called the Fisher information matrix is now familiar in papers dealing 
with the cosmological parameter estimation from CMB data, and expresses how 
much the likelihood is peaked around the best estimate.
Assuming the CMB and the noise fluctuations to be Gaussian, the Fisher 
information matrix is given by 
\beq
F_{ij}=\sum_l{2l+1\over 2}\left\{C_l+{4\pi\sigma^2\over N_p}
\exp\left[\theta_b^2 l(l+1)\right]\right\}^{-2}{\partial C_l\over \partial T_i}
{\partial C_l\over \partial T_j},
\eeq
where the additional exponential term comes from the beam smearing and 
its multiplicative coefficient from the instrumental noise. 

Let us summarize the 
limitations to the determination of the cosmological 
parameters from the CMB data. They can be classified in three categories: 
\begin{itemize}
\item Cosmic variance:
this comes from the fact that we are observing a limited number 
of  realizations 
of a random field. The cosmic variance is bigger at large angles and depends
on the sky coverage for a given experiment.

\item Degeneracies between cosmological parameters: 
one gets a degeneracy when different sets of cosmological parameters
reproduce the same anisotropy spectrum. A well-known example 
is the deneracy corresponding to a variation of $\Omega_k$ and 
$\Omega_\Lambda$ such that the angular size distance to the surface
of last scattering remains constant ($\Omega_bh^2$, $\Omega_ch^2$ fixed). 
This problem of degeneracies can in general be overcome by resorting 
to other types of cosmological observations, such as 
the supernova surveys or the large scale structure surveys. 

\item Degeneracies between cosmological parameters and the primordial 
spectra: this is the case when a variation of a cosmological parameter can 
be mimicked by a modification of the primordial fluctuation spectra.
\end{itemize}

Of course, additional  limitations come from the substraction of  the 
foregrounds from the CMB maps.

\section{What we hope to learn about the early universe 
(in the inflationary paradigm)}
Beyond the possibility of measuring  the cosmological parameters, 
 cross-checking with  other cosmological observations, the CMB
contains the extraordinary perspective to tell us something about the
early universe. 
One should be very cautious here, and not overstate  
the possibility to `see' early universe physics in the CMB. What the 
CMB will provide  is a consistency check for any
early universe model. Indeed, any early universe model, which explains
in an unambiguous way how the primordial fluctuations are generated, 
 can be confronted with the CMB observations, and either be rejected 
or be declared compatible with the data (with some constraints on the 
parameters of the model). However, what remains unknown to us 
is the amount of degeneracy 
among the early universe models themselves, that is how many consistent
early universe models can reproduce the same spectrum of cosmological 
perturbations. This applies in particular to the question of proving or
disproving inflation with the CMB data: the CMB data can tell us if the 
cosmological perturbations in the early universe follow a quasi-scale 
invariant spectrum; they cannot tell us directly 
if the actual mechanism which 
produced these fluctuations is really the gravitational amplification 
of the quantum fluctuations of a scalar field during a slow-roll phase. 
It is therefore always a healthy procedure in primordial 
cosmology to try to find alternative  models, based on different principles, 
with the condition of course that they must be  compatible with the data.

This being said, let us now review what the CMB could tell us, if we 
believe that inflation is the correct description of the early universe
(for more details, see the reviews by Mukhanov, Feldman and 
Brandenberger \cite{mfb92} and by Liddle and Lyth \cite{ll93}). 
The simplest models of inflation are based on a single scalar field, 
$\phi$, with a potential $V(\phi)$.
The phase of interest is the so-called slow rolling regime, when the 
scalar field is moving slowly, and which is 
characterized by the two conditions
\beq
\epsilon\equiv {m_P^2\over 2}\left({V'\over V}\right)^2 \ll 1, \quad
\eta\equiv m_P^2\left[{V''\over V}-{1\over 2}\left({V'\over V}\right)^2\right]
\ll 1, 
\eeq
where $m_P\equiv 1/\sqrt{8\pi G}$ is the reduced Planck mass. 
In the slow-roll regime, the homogeneous equations of motion 
for the scalar field reduce to 
\beq
3H\dot\phi\simeq -V', \quad 3H^2\simeq {V\over m_P^2}.
\eeq

During the inflationary phase, the scalar fluctuations of the metric 
can be written, in an appropriate coordinate system, in the form
\beq
ds^2=a^2\left[-(1+2\Phi)d\tau^2+\left(1-2\Phi\right)d{\vec x}^2\right].
\eeq
Defining  the correlation spectrum ${\cal P}_\Phi(k)$, 
in Fourier space,  by
\beq
\langle \Phi_{\vec k}\Phi^*_{\vec k'}\rangle=2\pi^2k^{-3}{\cal P}_\Phi(k)
\delta({\vec k}-{\vec k'}),
\eeq
the quantum fluctuations of the inflaton field, initially in 
their vacuum, can be shown to produce a fluctuation spectrum 
given by
\beq
{\cal P}_\Phi(k)=\left[\left({H^2\over\dot\phi^2}\right)
\left({H\over 2\pi}\right)^2\right]_{k=aH}
={128\pi\over 3}G^3\left({V^3\over V'^2}\right)_{k=aH},
\eeq
where the subscript means that the various quantities are evaluated 
at Hubble radius crossing, i.e. $k=aH$, for a scale $k$.
This spectrum (up to an overall constant according to the various 
definitions in the literature) is called the {\it scalar spectrum}. It is 
quasi-scale invariant since the scalar field, and thus the value 
of the potential and of its derivative, are supposed to vary 
slowly during this inflationary phase. 
One can also evaluate the scalar spectrum index $n_s(k)$ (which is 
weakly scale dependent for a slow roll regime):
\beq
n_s(k)-1\equiv{d\ln {\cal P}_\Phi(k)\over d\ln k}=2\eta- 4\epsilon.
\eeq

In addition to the scalar spectrum, the inflationary phase will 
also generate a spectrum of gravitational waves. 
Gravitational waves are perturbations of the metric of the 
form 
\beq 
ds^2= a^2\left[\eta_{\mu\nu}+h^{TT}_{\mu\nu}\right],
\eeq
where $h^{TT}_{\mu\nu}$ is a traceless transverse tensor. They possess only
two physical degrees of freedom (the two polarizations $+$ and $\times$), 
which  can be described effectively as two scalar fields 
\beq
\tilde \phi_{+,\times}={h_{+,\times}\over \sqrt{32\pi G}}.
\eeq
The spectrum of gravitational waves, or {\it tensor spectrum}, is then simply
 given by
\beq
{\cal P}_g(k)=2\times 32\pi G\times \left({H\over 2\pi}\right)^2.
\eeq
As for the scalar spectrum, one can define a tensor spectral index,
\beq
n_t(k)\equiv{d\ln {\cal P}_g(k)\over d\ln k}=-2\epsilon.
\eeq

An important consequence of the above results is that the ratio 
of the tensor and scalar amplitudes, ${\cal P}_g/{\cal P}_\Phi$ is 
proportional to $(V'/V)^2$, i.e. proportional to $\epsilon$. 
This implies that if one was  able to measure the scalar amplitude as well 
as the tensor amplitude and tensor index, then 
 one could {\it check}  if this  consistency
relation is satisfied. This would probably represent 
 one of the most significant tests for inflation.

After this general introduction to inflation, 
let us  just present the three  main  categories of models 
(many more details on the numerous models of inflation and their link 
with particle physics  
 can be found in a recent review by Lyth and Riotto \cite{lr99}) .

\subsection{Chaotic type models ($0<\eta\leq \epsilon$)}
This category corresponds to models 
 with a scalar field amplitude of the order of a few $m_P$
during slow roll inflation and with a potential typically of the form
\beq
V(\phi)=\Lambda^4\left({\phi\over\mu}\right)^p.
\eeq
These extremely simple potentials for inflation have been initially introduced
in the context of so-called `chaotic inflation'. 
Another typical potential in this category is the exponential 
potential,
\beq
 V(\phi)=\Lambda^4\exp\left(\phi/\mu\right),
\eeq
which gives rise to power-law inflation, where  the scale 
factor evolves like $a(t)\propto t^{1/\epsilon}$, 
with 
$
\epsilon=\eta=\left({m_P/ \mu}\right)^2/2$.

This category of models has had a lot of success in the literature because 
of their computational simplicity. However, in general, they are not 
considered to be models which can be well motivated by   particle 
physics. The reason is the following. The generic potential 
for a scalar field will contain an infinite number of terms,
\beq
V(\phi)=V_0+{1\over 2}m^2\phi^2+{\lambda_3\over 3}\phi^3
+{\lambda_4\over 4}\phi^4+\sum_{d=5}^\infty m_P^{4-d}\phi^d,
\eeq
where the non-renormalizable  ($d>4$) couplings $\lambda_d$ are a priori 
of order $1$. When the scalar field is of order of a
few Planck masses,  one has no control on the form of the 
potential, and  all the non-normalizable terms must be taken into 
account in principle. In order to work with more specific forms for the 
potential,  inflationary model builders tend to 
concentrate on models where the scalar field amplitude is small with 
respect to the Planck mass.

\subsection{Spontaneous symmetry breaking  models ($\eta<0<\epsilon $)}
This type of models is  characterized 
by $V''(\phi)<0$, and 
a typical potential can be written in the form 
\beq
V(\phi)=\Lambda^4 \left[1-\left({\phi\over\mu}\right)^p\right].
\eeq
This  can be interpreted as the lowest-order term in a Taylor expansion 
around the origin. Historically, this potential shape appeared in 
the so-called `new inflation' scenario, which followed the 
initial  version of inflation, dubbed `old inflation', based on a 
first-order transition which was shown to never end at least if one 
wants enough inflation to solve the usual horizon problem of 
standard cosmology. 

A particular feature of these  models is that tensor modes 
are much more  suppressed with respect to scalar modes than in the 
large-field
models. 

\subsection{Hybrid models ($0<\epsilon<\eta $)}
A new category of models has appeared more recently. In the hybrid 
scenario, 
two scalar fields are taken into account. One scalar field is responsible 
for inflation and evolves toward a minimum with nonzero vacuum energy,
 while the second field is responsible for the end of inflation. 
The `decoupling' between the inflaton and the end of inflation allows 
a richer range of possibilities. 

Hybrid inflation potentials, 
which frequently appear in supersymmetric models,
 are characterized by $V''(\phi)>0$ and $0<\epsilon<\eta$.
As far as the spectrum of perturbations is considered, one needs only the 
shape of the effective potential of the inflaton field during inflation, 
which can be described in the form
\beq
 V(\phi)=\Lambda^4 \left[1+\left({\phi\over\mu}\right)^p\right].
\eeq
Once more, this potential can be seen as the lowest order in a Taylor 
expansion around the origin. 
The value $\phi_N$ of the scalar field as a function of the number of 
e-folds before the end of inflation is not determined by the above potential
and, therefore, $(\phi_N/\mu)$ can be considered as a freely adjustable 
parameter. 
A characteristic feature of hybrid models is that they can lead, in 
contrast to the two previous categories,  to a
blue spectrum ($n_s>1$), although there are also models of hybrid 
inflation giving a red spectrum.

\section{The importance of CMB polarization}
Although this section is somehow related to the previous one, it may be 
useful 
to devote a special section to the subject of the CMB polarization, 
the measurement of which might open a qualitative new window on the early 
universe. Before discussing its connection with  the primordial universe, 
in particular within the inflationary paradigm, let us first review 
the basic formalism used to describe the polarization \cite{polarization}.

The linear polarisation of the CMB can be described by a (two-index) 
tensor field  on the two-sphere, in the same way as the CMB temperature is 
described by a scalar field on the two sphere. 
This tensor field can be decomposed onto a basis of  tensor spherical 
harmonics, themselves separable into the  electric-type tensor 
harmonics
\beq
Y^E_{(lm)ab}=N_l\left(Y_{(lm);ab}-{1\over 2}\gamma_{ab}Y_{(lm);c}^{\, \, ;c}
\right),
\eeq
and the magnetic-type tensor harmonics
\beq
Y^B_{(lm)ab}={N_l\over 2}\left(Y_{(lm);ac}\varepsilon^c_{\, b}
+Y_{(lm);bc}\varepsilon^c_{\, a}
\right),
\eeq
with the normalisation constant $N_l=\sqrt{2(l-2)!/(l+2)!}\, $, and 
where $\gamma_{ab}$ is the metric on the two-sphere and 
$\varepsilon_{ab}$ the associated antisymmetric tensor.
In analogy with the multipole coefficients for the temperature defined in 
(\ref{alm}), which will be now denoted $a_{lm}^T$, one can define the 
electric multipole coefficients $a^E_{lm}$ and the magnetic 
multipole coefficients $a^B_{lm}$. As a consequence, there will be 
in total four angular power spectra. In addition to the temperature 
power spectrum $C_l^T$ defined in (\ref{Cl}), one finds
the electric and magnetic power spectra,
\beq
C_l^E=\langle |a^E_{lm}|^2\rangle, \quad
C_l^B=\langle |a^B_{lm}|^2\rangle,
\eeq
and the cross-correlation spectrum
\beq
C_l^C=\langle a^T_{lm} a^{E*}_{lm}\rangle.
\eeq
The two other cross correlations one could envisage, $E$ with $B$ or 
$T$ with $B$, automatically vanish for reasons of symmetry.

 The importance of CMB polarization concerning the physics of the early 
universe relies on the fact that scalar fluctuations can produce only
E-type polarization and no B-type polarization. A measurement of the 
B-polarization  could therefore be interpreted as the detection of 
primordial gravitational waves.
There are unfortunately  a few problems in this attractive perspective. 
Obviously, the first difficulty is the smallness of the signal: the CMB is 
expected to be polarized only at the $5-10 \%$ level on small angular 
scales, 
and even less on large angular scales. Detection of the CMB polarization is 
thus in itself a technological challenge. 
Another problem concerns the interpretation of a positive signal: not 
only primordial gravitational waves but also foreground emission, or 
any process with Faraday rotation, will generate B-type polarization. In 
principle, these contaminants could be substracted upon use of 
multi-frequency observations, but this makes the objective of actually 
measuring the amount of primordial waves still more remote.

\section{More exotic possibilities}
So far, we have presented the most consensual picture for the 
early universe with its relation to actual observations. In this sense, 
the slow-rolling single field inflationary scenario represents today 
the {\it minimal standard model of the early universe}. Within this 
perspective, the future path of research is well paved: with the 
constant refinement of CMB measurements expected for the forthcoming years, 
the constraints on the various parameters describing the slow-roll 
regime  (and even beyond the slow-roll approximation) 
will be tighter and tighter, thus excluding more and more 
inflationary scenarios. 

Of course, the early universe cosmologists have been trying constantly 
to explore alternative avenues. A long-standing opponent to the 
inflationary paradigm has been the formation of structure seeded by 
topological defects, which are predicted to be produced in Grand Unified 
Theories. The recent CMB data have shown that 
these models cannot explain, by themselves, the observed signal. 
Another interesting idea is the pre-big-bang scenario \cite{prebigbang}, 
trying to make the connection between string theory and the early universe.
This scenario unfortunately suffers from two weaknesses: the transition 
between the pre-big-bang and the post-big-bang, where the physics is not 
under control so far; and the fact that it does not seem to give  
 a nearly scale-invariant 
scalar fluctuation spectrum. 

In the exploration of more exotic scenarios, one must distinguish two 
types of approaches: one consists in considering more refined versions 
of the simplest model, thus adding more degrees of freedom 
 than usually necessary to fit the data.  
This attitude is useful for two reasons: for exploring 
the robustness of the main model (i.e. this is in some sense a study 
of the degeneracy among early universe models in the neighbourhood  of 
the reference model); for studying the possibilities of obtaining 
from the cosmological data some extra information which simply does not 
exist  
in the main model. 
Another, more radical approach, consists in trying to find completely new 
types of scenario, and see if they can reproduce the impressive successes 
of the inflationary scenario. 
We will now give one  example for each of  these two approaches. 

\subsection{Isocurvature perturbations}
In the single field inflationary picture, the cosmological fluctuations 
produced during inflation are necessarily {\it adiabatic}, i.e. the 
 relative composition  in the various cosmological  species is the 
same for  the pertubations as for  the background, because all species 
have the same origin: the unique scalar field. 
One must however keep in mind the possibility of {\it isocurvature}
perturbations, defined  in the case of two species 
$X$ and $Y$ by the non-vanishing quantity
\beq
S_{X,Y}={\delta n_X\over n_X}-{\delta n_Y\over n_Y},
\eeq
where $n_{X,Y}$ are  the number densities of the species, the 
perturbation in the total energy density being zero.  
Although pure isocurvature primordial spectra have been shown to be 
incompatible with cosmological data, a small fraction of 
isocurvature primordial perturbations is allowed although strongly 
constrained.

What has recently renewed the interest  in isocurvature perturbations
is the richer range of possibilities 
 if these isocurvature perturbations are {\it correlated} 
with the usual adiabatic perturbations. This correlation 
has been first illustrated \cite{langlois99} in 
  a very simple model of double inflation with two free massive scalar 
fields, with the Lagrangian
\beq
{\cal L}=-\partial_\mu \phi_h\partial^\mu \phi_1 -{1\over 2}m_1^2\phi_1^2
-\partial_\mu \phi_l\partial^\mu \phi_2 -{1\over 2}m_2^2\phi_2^2.
\eeq
The two scalar fields being uncoupled, their quantum fluctuations 
$\delta\phi_1$ and $\delta\phi_2$ are statistically independent. However
both fields will in general contribute to the primordial 
adiabatic and isocurvature perturbations,
\beq
\Phi=A_1\delta\phi_1 + A_2\delta\phi_2, \quad
S=B_1\delta\phi_1 + B_2\delta\phi_2,
\eeq
where the $A$'s and $B$'s are background dependent coefficients. 
It turns out that in some region of the parameter space, $\Phi$ and 
$S$ will be correlated. Allowing for correlation between isocurvature 
and adiabatic perturbations gives more freedom to play with the predicted 
CMB spectrum and has been more systematically studied in several 
recent works \cite{correlated}.

\subsection{Brane cosmology}
The idea of extra-dimensions has recently gone through a renewal with
 the 
hypothesis, suggested by recent developments in string theory, that ordinary
matter is confined to a three-dimensional subspace, or {\it brane}, 
embedded in a higher 
dimensional spacetime or {\it bulk}. In cosmology, particular attention 
has been devoted to five-dimensional models where the worldsheet of our 
 Universe-brane  is  a hypersurface.

A striking result, obtained  when solving 
the  five-dimensional Einstein's equations $G_{AB}\equiv R_{AB}-R g_{AB}/2=
\kappa^2 T_{AB}$,  is that the matter content of 
the brane enters {\it quadratically} \cite{bdl99}
 in the Friedmann  equations  instead of linearly as in standard cosmology.
In the case of an empty bulk, one would thus find a cosmological evolution 
incompatible with our understanding of nucleosynthesis.
 A way out has been 
found by applying the Randall-Sundrum idea \cite{rs99b} 
to cosmology \cite{cosmors,bdel99}, i.e. considering
an Anti-de Sitter  bulk spacetime (with 
a negative cosmological constant $\Lambda$) and a tension in the brane.  
The (assumed) cancellation of $\Lambda$ with the square of the brane tension 
$\sigma$
leads to the new Friedmann equations \cite{kraus,bdel99}
\beq
H^2= {8\pi G\over 3}\left(\rho+{\rho^2\over 2\sigma}\right),
\label{new_friedmann}
\eeq
where $H$ is the Hubble parameter in the brane, $\rho$ the cosmological energy
density in the brane. And Newton's constant is related to the brane tension 
by $8\pi G=\kappa^4\sigma/6$.
This equation gives the usual evolution in the
low energy regime $\rho\ll\sigma$ and quadratic corrections in the 
high energy regime $\rho > \sigma$.

The next step is obviously  the influence of 
extra-dimensions on the {\it cosmological perturbations} and their evolution, 
and thus try to make the link with a specific signature for the CMB 
predictions.  
Several pioneering works have developed formalisms to handle the cosmological 
perturbations for a brane-universe in a five-dimensional spacetime 
(see \cite{l00B} and references therein).
In the case of slow-roll inflation generated by a scalar field 
confined to the brane \cite{mwbh99}, one can compute 
explicitly the  cosmological fluctuations generated during a quasi-
de Sitter phase, both for the scalar spectrum \cite{mwbh99} and the 
tensor spectrum \cite{lmw00}. However, for the subsequent 
 radiation and matter dominated eras, the evolution of perturbations is much
more complicated, except in the case of super-Hubble perturbations 
\cite{lmsw00},  because  a quantitative 
analysis and therefore the determination of the CMB anisotropy spectrum 
depends on the specific 
distribution of gravitational waves in the bulk.

\section{Conclusions}
It is always extremely difficult to make predictions about the future of 
a scientific domain and I will not take this risk. 
The amount of information we will learn from 
the future CMB observations will depend on how close or how far they 
turn out to be with respect to the {\it canonical} version of the early 
universe model, that of an inflationary phase generated by a single 
field in slow-roll motion. 
To be too close or too far are probably the cases where the gain of 
information will be  minimal for theorists. 
Obviously, being very close would be  a tremendous success for 
the currently prefered model but would not bring any surprise. 
However, 
being too far might  not provide so much information as well, unless 
it corresponds to the predictions of a model already considered.
We would have to revise some of our ideas but it is 
usually  easy for theorists to cook up a model, or even several, 
which will fit the data {\it a posteriori}.
The most stimulating situation 
  for cosmology, where the number of observations concerning
the early universe is extremely limited, would occur if  the new data roughly 
confirm the overall picture but add     details that 
reveal some deviations from 
the canonical model. This is the best situation to learn something because
this is the case where one is most likely to interpret correctly 
the unexpected features of the data.
Let us hope therefore than the next observations will put us in such 
a position.

\ack
It is a pleasure  to thank the organizers for this remarkable conference. 
I would also like to thank the {\it Ambassade de France} in South Africa for 
their help and financial support.  



\end{document}